# Cost-Effective Community-Hierarchy-Based Mutual Voting Approach for Influence Maximization in Complex Networks

Yi Liu, Xiaoan Tang, Witold Pedrycz, *Life Fellow*, *IEEE*, Qiang Zhang

*Abstract*—Various types of promising techniques have come into being for influence maximization whose aim is to identify influential nodes in complex networks. In essence, real-world applications usually have high requirements on the balance between time complexity and accuracy of influential nodes identification. To address the challenges of imperfect node influence measurement and inefficient seed nodes selection mechanism in such class of foregoing techniques, this article proposes a novel approach called Cost-Effective Community-Hierarchy-Based Mutual Voting for influence maximization in complex networks. First, we develop a method for measuring the importance of different nodes in networks based on an original concept of Dual-Scale Community-Hierarchy Information that synthesizes both hierarchy structural information and community structural information of nodes. The community structural information contained in the nodes is measured by a new notion of Hierarchical-Community Entropy. Second, we develop a method named Cost-Effective Mutual-Influence-based Voting for seed nodes selection. Hereinto, a low-computational-cost mutual voting mechanism and an updating strategy called Lazy Score Updating Strategy are newly constructed for optimizing the selecting of seed nodes. Third, we develop a balance index to evaluate the performance of different methods in striking the tradeoff between time complexity and the accuracy of influential nodes identification. Finally, we demonstrate the approach performance over ten public datasets. The extensive experiments show that the proposed approach outperforms 16 state-of-the-art techniques on the balance between time complexity and accuracy of influential nodes identification. Compared with the method with the second highest value of the balance index, our approach can be improved by at most 9.29%.

This work was supported in part by the Natural Science Foundation of China (Nos. 72101075, 72171069, and 92367206), in part by the Natural Science Foundation of Anhui Province of China (No. 2108085QG289), and in part by the Key laboratory of Industrial Equipment Quality Big Data, MIIT (No. 2023-IEQBD-02). *(Corresponding author: Xiaoan Tang.)*

Yi Liu, Xiaoan Tang and Qiang Zhang are with the Key Laboratory of Process Optimization and Intelligent Decision-making, Ministry of Education, the School of Management, Hefei University of Technology, Hefei, Box 270, Hefei 230009, Anhui, P.R. China (e-mails: 2019210381@mail.hfut.edu.cn; tangxa@hfut.edu.cn; qiang_zhang@hfut.edu.cn).

Witold Pedrycz is with the Department of Electrical and Computer Engineering, University of Alberta, Edmonton, AB T6G 2R3, Canada, also with the Systems Research Institute, Polish Academy of Sciences, 00-901 Warsaw, Poland, Istinye University, Faculty of Engineering and Natural Sciences, Department of Computer Engineering, Sariyer/Istanbul, Turkiye (e-mail: wpedrycz@ualberta.ca).

## I. INTRODUCTION

Commonly encountered complex networks as transmission networks [1], social networks [2] and internet of vehicles [3] have gained great attention of many researches [4-6] due to their ability to be used to model real-world interconnections among various objects. Over the past few decades, researchers have studied the structures, characteristics, and other aspects of these complex networks, resulting in lots of interesting research findings that involve community detection [7], influence maximization [8], among others. In particular, influence maximization has become a popular research topic due to its practical applications in the fields of crucial criminal identification [9], target advertising [10], viral marketing [11], rumors suppression [12, 13], and emergency information diffusion [14].

The influence maximization problem is to find a set of seed nodes that can generate maximum influence spread in complex networks. Over the years, various types of promising techniques like greedy-based algorithms, heuristic algorithms, and community-structure-based algorithms have been proposed for influence maximization and they demonstrated distinctive performance in terms of time complexity and accuracy of influential nodes identification. Kempe *et al.* [15] proved that obtaining an optimal solution to the influence maximization problem is an NP-hard optimization problem and developed the original greedy-based algorithm to solve the problem. Next, Leskovec *et al.* [16] proposed the Cost-Efficient Lazy Forward (CELF) algorithm based on the sub-modularity of the influence expectation function, which can greatly improve the efficiency of the original one. Goyal *et al.* [17] further proposed the CELF++ algorithm, which outperforms the CELF algorithm. The more nodes the seed nodes found by one method can influence, the better the method performs in the accuracy of influential nodes identification. Well as they perform in the accuracy of influential nodes identification, these greedy-based methods have attracted little attention in recent years due to their unacceptable time complexity in large-scale complex networks caused by generating a large amount of simulation processes.

As such, researchers have endeavored to use structural features or specific attributes of the nodes to replace the time-consuming simulation processes in order to filter out influential nodes in the networks, which gives rise to the development of heuristic methods. On the one hand, some heuristic methods like the centrality-based algorithms (e.g., the k-shell method [18], the H-index-based methods [19, 20], the Locality-based Structure System (LSS) [21], and the Extended Cluster Coefficient Ranking Measure (ECRM) approach [22]) perform well on reducing the time complexity when being used to identify influential or important nodes in the networks. Inspired by Shannon's information entropy, Zhang *et al.* [23] proposed the Local Fuzzy Information Centrality (LFIC). Zhang *et al.* [24] proposed the Laplacian Gravity Centrality (LGC) method which can successfully address the problem of original gravity centrality to identify influential nodes. Zhao *et al.* [25] proposed the SHKS method by combining k-shell and structural holes, which can successfully recognize some nodes with small k-shell indices but good propagation ability. These centrality-based methods, although effective in reducing the time complexity compared to the greedy-based methods, suffer from the problems of over-concentration of seed nodes and overlapping influence of different seed nodes. On the other hand, researchers have found this issue and proposed many improved heuristic methods. Gupta *et al.* [26] proposed the Relative Local-Global Importance (RLGI) method considering the feature that nodes located in the same core may be removed at different rounds in the k-core decomposition and combining the local and global information for nodes. Zhang *et al.* [27] utilized the voting mechanism to select seed nodes and adopted certain suppression measures to solve the problem of overlapping influence. Liu *et al.* proposed the VoteRank++ algorithm [28] and the VoteRank* algorithm [29] by giving nodes different initial voting abilities according to their importance and taking further suppression measures for the second-order neighbors of seed nodes, etc. These improved heuristic methods, although they mitigate the problem of overly concentrated distribution of seed nodes to a certain extent, still have certain shortcomings in terms of accuracy because they do not adequately consider the conditions required for information to spread over a wide range of the networks. And the time complexity of certain methods of this kind is still high due to the related suppression measures and corresponding redundant calculations.

In order to disseminate information as widely as possible in the networks, researchers have set their sights on the community structures of complex networks. They argued that the characteristics of closer ties within the community than outside make it easier for information to be disseminated within the community, and the process of information dissemination on a large scale in the networks is the process of information circulation between different communities. For this reason, by identifying the core nodes of different communities as influential nodes, the community-structure-based methods like the Community-based k-shell decomposition (CKS) algorithm [30] and the Community-based Influence Maximization (CIM) algorithm [31] have been proposed. More specifically, Yang *et al.* [31] classified the nodes into three types of peaks, slopes, and valleys by the concept of topological potential field and finally extracted core nodes in the important communities. Kazemzadeh *et al.* [32] proposed the method of Charismatic Transmission in Influence Maximization (CTIM), which prunes unsuitable communities to reduce computational cost and uses global diffusion power to select the most influential nodes among different communities. Umrawal *et al.* [33] developed a generic community-aware framework for influence maximization. At the same time, certain researchers argued that the practice of considering only the core nodes of different neighborhoods may allow information about community-to-community connectivity to be ignored. Focusing on bridge nodes between communities, Kumar *et al.* [34] designed the Community-based Spreaders Ranking (CSR) method in terms of community diversity, number of nodes contained in the community, and density of the community. Subsequently, Kumar *et al.* [35] proposed the Community-structure with Integrated Features Ranking (CIFR) algorithm by integrating the key nodes within the community as well as the gateway nodes of the community and using the number of links between different communities to measure the importance of the communities. Li *et al.* [36] introduced the Layered Gravity Bridge (LGB) algorithm, which utilizes local betweenness centrality, external connection metrics as well as the community merging process to find the core nodes within the community and the bridge nodes between the communities. These community-structure-based methods effectively solve the problem of influence overlapping in the networks to a certain extent while maintaining relatively low complexity due to the replacement of global network-based computation with local community-based computation. They can strike a relative balance between time complexity and the accuracy of influential nodes identification. These methods can address the network-level influence overlapping problem to some extent, while they also raise the influence overlapping problem at the community-subgraph level, especially when the community subgraphs are large. Some of the methods delete certain unimportant communities in order to reduce the time complexity, which may lead to the loss of some of the network information.

Many real-world marketing applications usually have high requirements on the balance between time complexity and accuracy of influential nodes identification. For example, in commercial marketing campaigns, we may not be able to pay the compensation demanded by the biggest influencers we find due to budgetary constraints [37, 38]. Meanwhile, we are only able to conduct a limited number of marketing campaigns within a given time frame [38] because too many marketing campaigns may cause consumer fatigue [39] and each round of marketing campaigns requires us to engage in commercial negotiations with the influencers [40], which obviously takes up a certain amount of time. In this case, how to effectively balance time complexity and accuracy of influential nodes identification turns out to be a critical and interesting issue [38, 41, 42].

Motivated by the above ideas, this study proposes a computational cost-effective approach called Cost-Effective Community-Hierarchy-Based Mutual Voting (CECHMV) for solving the influence maximization problem while keeping a good balance between time complexity and the accuracy of influential nodes identification. The overall approach framework contains the development of a Dual-Scale Community-Hierarchy Information (DSCHI) method for node importance measurement and the presentation of a technique called Cost-Effective Mutual-Influence-based Voting (CEMIV) used to seed nodes selection. The main contributions of this article are outlined as follows:

1) We develop a method for measuring the importance of

different nodes in networks based on a newly-introduced concept of DSCHI that synthesizes both hierarchy structural information and community structural information of nodes. Hereinto, the community structural information contained in the nodes is measured by a new notion of Hierarchical-Community Entropy (HCE).

2) We develop a method named CEMIV for seed nodes selection with a wonderful balance between time complexity and the accuracy of influential nodes identification, inspired by real-world voting scenarios and based on the aforementioned DSCHI. Within the developed method, a low-computational-cost mutual voting mechanism and an updating strategy called Lazy Score Updating Strategy (LSUS) that exhibits sound generalization are newly constructed for optimizing the selecting of seed nodes.

3) We originally develop a *Balance Index* by integrating the commonly used evaluation criteria of *Final Infected Scale* and *Running Time* to quantify and measure the performance of different methods in striking a tradeoff between time complexity and the accuracy of influential nodes identification.

4) We demonstrate the approach performance over ten public available complex network datasets. The extensive experiments show that the proposed CECHMV approach outperforms 16 state-of-the-art techniques on the balance between time complexity and accuracy of influential nodes identification when solving influence maximization problem. Compared with the method with the second highest value of the *Balance Index*, our method can be improved by at most 9.29%.

The originality of this study is fourfold: the introduction of the DSCHI method, the development of the CEMIV method, the design of the LSUS, and the presentation of the *Balance Index*. The paper is organized as follows. Section II presents some basic concepts and classical models related to influence maximization. Section III describes the details of the proposed CECHMV approach consisting of the algorithms of DSCHI and CEMIV used to effectively and efficiently solving influence maximization problems. Section IV covers the experimental results in comparison with 16 state-of-the-art methods on ten datasets to demonstrate the performance of the proposed approach. Finally, Section V concludes this article and identifies possible future research directions.

## II. PRELIMINARIES

### A. Basic Concepts and Problem Definition

A complex network is usually denoted by $G(V, E)$, where $V$ and $E$ denote the set of nodes and the set of edges in the network, respectively. $|V|$ and $|E|$ denote the corresponding numbers of nodes and edges and $<D> = 2|E|/|V|$ denotes the average degree of the network. As the average degree increases, the number of connections between nodes also increases [43, 44]. The influence maximization problem can be defined as finding a seed set $S_\sigma \subseteq V$ of $\varphi$ nodes that can influence the maximum number of nodes in the network, where $\varphi$ is a positive integer and $\varphi < |V|$ [33]. The problem can be formally defined as [35]:

$$S_\sigma = argmax(\delta(S)) \quad (1)$$

where $\delta(S)$ denotes the number of nodes that the seed nodes set $S$ can influence in the given network, and $S$ denotes any set of $\varphi$ seed nodes in $V$.

A community is a subgraph of the global complex network graph, where the nodes within the same community are closely connected, and the nodes between different communities are sparsely connected. The issue of how to mine latent community structures in complex network is an open research topic and lots of methods have been reported in the field of community detection [7, 45, 46].

### B. Susceptible-Infected-Recovered model

The Susceptible-Infected-Recovered (SIR) model [47] can be used to effectively simulate the diffusion process of certain kinds of diseases. Since the similarities between the influence propagation process in the network and the spreading process of infectious diseases, the SIR model is a commonly used tool for simulating the influence propagation process and measuring the influence of nodes. The nodes in the network are classified into three categories, namely Susceptible (S), Infected (I), and Recovered (R). Specifically, the nodes in the S state are currently not infected but able to be infected by their neighbor nodes. The nodes in the I state are currently infected and can infect their neighbor nodes. The nodes in the R state are recovered from the infection and will not be infected again. During the propagation process, each node in State I will try to infect its neighbor nodes which are in State S with probability $\mu$ and will attempt to recover to State R with probability $\beta$. The work done in [27] shows that the infected rate $\lambda = u/\beta$ plays a significant role in affecting the speed of propagation and the final size of the propagation.

### C. K-shell Method and Neighborhood Coreness

Kitsak *et al.* [18] argued that nodes at the core position of a network tend to be more influential than those with high connectivity and introduced the k-shell method. They first iteratively remove nodes of degree 1 from the graph, thereby dividing the different shells and assigning each node a shell value, and then they use different shell values to differentiate the importance of different nodes. But there are often situations where a large number of nodes have the same shell value. Bae *et al.* [48] attempted to use the sum of shell value of a nodes' neighbor nodes as a novel index to measure the importance of a node and therefore proposed the neighborhood coreness (NC) method to address these situations.

## III. PROPOSED METHOD

In this section, we present the CECHMV approach in detail. First, the DSCHI method for measuring the importance of different nodes in networks is introduced by taking both the condition required for information to spread over the global networks, which is the community, and the hierarchical characteristic of nodes into consideration. Then, the CEMIV method for seed nodes selection is presented inspired by real-world voting scenarios and based on the aforementioned

DSCHI method. Finally, the whole algorithmic framework of the proposed CECHMV approach is presented.

*A. Dual-Scale Community-Hierarchy Information*

In a network, each community may be of different importance for the reasons of their different node numbers and density, positions, etc. Especially, communities located at the core of the network are often more effective in disseminating information. As such, the position information of the communities which is always ignored in most existing studies should be considered when we are to measure the importance of the communities. To do this, we first introduce one concept of Community Coreness (CC) to measure the degree of coregency of a community's position in the network.

Given a graph $G_c(V_c, E_c)$, which is not a multigraph [35] and consists of different communities, where $V_c$ is the set of points of $G_c$, i.e., different communities, $E_c$ denotes the set of edges of $G_c$, i.e., edges between different communities. Executing the k-shell algorithm [18] once on $G_c$, and as for the node $i \in V_c$, we can obtain its $ks$ value:

$$ks(i) = k - shell(G_c(V_c, E_c)) \quad (2)$$

and then, the CC value of node $i$ can be computed as follows:

$$CC(i) = \sum_{w \in Neighbor(i)} ks(w) \quad (3)$$

where $Neighbor(i)$ denotes the set of neighbor nodes of node $i$. By normalizing the CC value of all nodes, we have:

$$CC_N(i) = \frac{CC(i)}{CC_{max}} \quad (4)$$

where $CC_{max}$ is the maximum CC value of all nodes.

Considering that nodes at the crossroads of different communities tend to be more powerful in disseminating information from one community to another. Inspired by the information entropy [49], we then introduce the other concept of Hierarchical-Community Entropy (HCE) to measure the amount of the community structural information of nodes. For the $m$ communities, the Community Importance (CI) of each community can be calculated as follows:

$$CI(C_l) = NN(C_l) * CC_N(C_l) \quad (5)$$

where $NN(C_l)$ is the number of nodes in the $l$th community $C_l$. Then, we can calculate the HCE of each node $v$ in the global network $G$ in the following manner:

$$HCE(v) = -\left(\frac{1}{2}\sum_{l=1,l \neq z}^{m} \frac{1}{(CI_{max}-CI(C_l)+1)} * \frac{ne_{C_l}}{d(v)} * log_2\left(\frac{ne_{C_l}}{d(v)}\right)\right. \\ \left. + \frac{1}{(CI_{max}-CI(C_z)+1)} * \frac{ne_{C_z}}{d(v)} * log_2\left(\frac{ne_{C_z}}{d(v)}\right)\right) \quad (6)$$

where $ne_{C_l}$ denotes the number of neighbor nodes of node $v$ in community $C_l$ that does not include the community in which node $v$ itself is located. $C_z$ denotes the community in which node $v$ itself is located and $ne_{C_z}$ denotes the number of neighbor nodes of node $v$ in community $C_z$. $CI_{max}$ denotes the maximum CI value of all communities and $d(v)$ denotes the degree of node $v$.

By normalizing the HCE of all nodes $v$, we have:

$$HCE_N(v) = log_2\left(1 + \frac{HCE(v)}{HCE_{max}}\right) \quad (7)$$

where $HCE_{max}$ denotes the maximum value of the HCE values of all nodes.

The following aspects are taken into account when measuring the amount of community structural information a node contains: the number of communities it connects to, the diversity of the communities it connects to, the importance of the communities it connects to, and the extent of its own influence on different communities. If a node and all of its neighbor nodes lie in the same community, the node has no role in information dissemination across communities, so in this case we argue that the node contains no community structural information. It is clear that nodes at the intersection of multiple communities tend to be better able to influence the communities to which they belong, but this is often overlooked by most of existing studies. Different from the coarse-grained methods [34-36] in selecting the bridge nodes between communities, using the concept of HCE can effectively evaluate the amount of community structural information of the nodes contain in a fine-grained level without needing to delete any communities. A detailed demonstration example of the process of calculating the HCE value is presented in Section A.1 of Appendix A of the supplementary material.

For-as-much-as nodes at the core position of a network tend to be more influential [18], so we also need to consider the degree of core of a node's position in the network when considering its influence. As such, in what follows, we propose the concept of Dual-Scale Community-Hierarchy Information (DSCHI) to synthesize a node's community structural information as well as its hierarchy structural information to effectively measure the importance of a node.

Given a global graph $G(V, E)$, where $V$ is the set of points of $G$, $E$ denotes the set of edges of $G$, similarly, executing the k-shell algorithm [18] once on $G$, and as for the node $v \in V$, we can obtain its $ks$ value:

$$ks(v) = k - shell(G(V, E)) \quad (8)$$

Based on the NC method proposed by Bae et al. [44], we can obtain the NC value of node $v$:

$$NC(v) = \sum_{u \in Neighbor(v)} ks(u) \quad (9)$$

where $Neighbor(v)$ denotes the set of neighbor nodes of node $v$. By normalizing the NC value of all nodes, we have:

$$NC_N(v) = log_2\left(1 + \frac{NC(v)}{NC_{max}}\right) \quad (10)$$

where $NC_{max}$ denotes the maximum value of the NC values of all nodes. Then, we can calculate the DSCHI value of each node $v$ in the following manner:

$$DSCHI(v) = \alpha * HCE_N(v) + (1 - \alpha) * NC_N(v) \quad (11)$$

where $\alpha$ is an adjustment coefficient and $\alpha \in (0,1)$. By adjusting the $\alpha$, we can control the weight of the community structural information and hierarchy structural information in measuring the importance of the nodes. In order to achieve broader information dissemination in a network, it is necessary to prioritize the role of nodes as bridges between different regions of the network, as such we set $\alpha$ to 0.7 in this study.

It is obvious that the development of DSCHI for node importance measurement comprehensively considers the degree of core of node position in the network and its bridge role in information dissemination across communities. As a whole, below we present the method framework in Fig.1, the algorithmic representation in Algorithm 1 and detail steps of the proposed DSCHI method.

*Step 1:* In view of the outstanding performance of the Leiden algorithm [46] in community detection in networks of different scales, by which we detect the community structure of the network. After acquiring the communities, we calculate

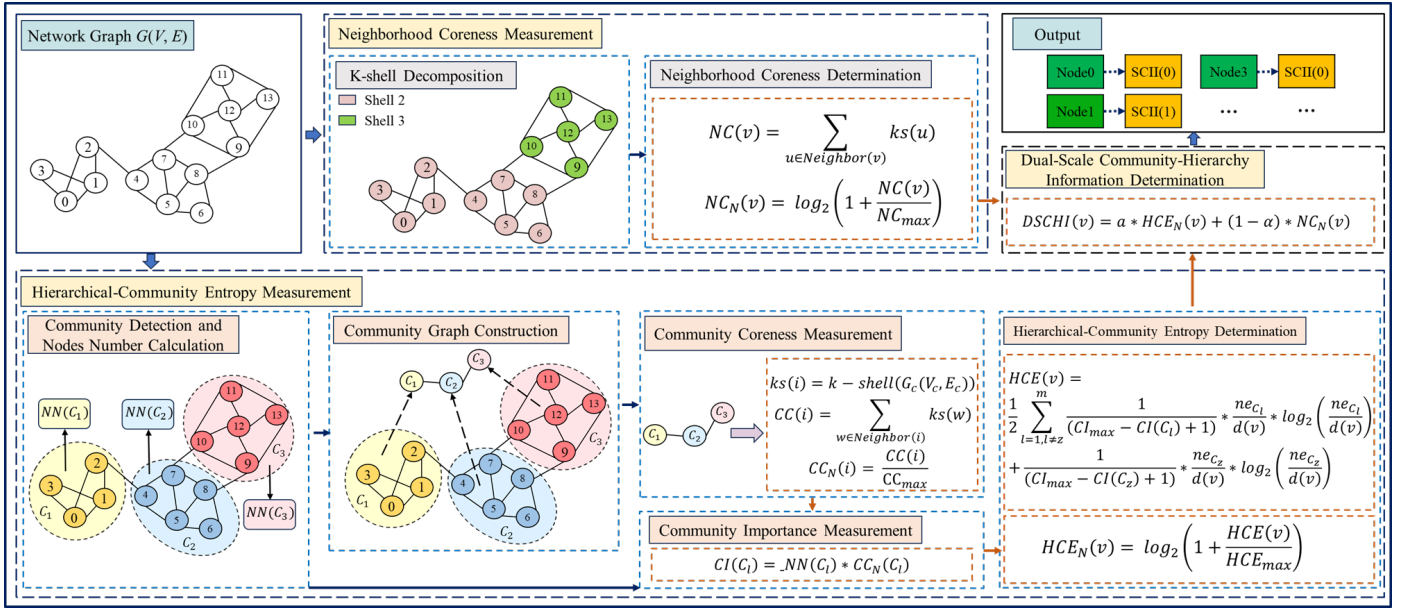

**Fig. 1.** The framework of the proposed DSCHI method

the number of nodes of each community and determine the NN value of each community.

*Step 2:* In this step, we execute a round of traversal of the nodes in the network. During the process, we construct the community graph $G_c$ and acquire the number of neighbor nodes of each node in each of its connected communities. After acquiring the community graph $G_c$, we can calculate the CI value of each community by using Equations (2)-(5).

*Step 3:* In this step, we execute two rounds of traversal of the nodes in the network. In the first round, we acquire the HCE value of each node by using Equations (6) and (7). In the second round, we can calculate the NC value of each node by using Equations (8)-(10). We then can get the DSCHI value of each node in the network.

---

Algorithm 1: DSCHI Algorithm Framework
**Input**: $G(V,E)$
**Output**: A Dict consisting of different nodes and corresponding DSCHI value for different nodes
1  $C \leftarrow$ Use Leiden Algorithm to detect communities;
2  **foreach** community $C_l \in C$ **do**
3    $NN(C_l) \leftarrow$ compute the number of nodes of each community;
4  **foreach** node $v \in G$ **do**
5    construct the graph $G_c$ of different communities;
6    **foreach** node $l \in G_c$ **do**
7      compute $ne_{C_l}$;
8  **foreach** node $i \in G_c$ **do**
9    compute $CC_N(i)$ using (2), (3), (4) in turn;
10   compute $CI(i)$ using (5);
11 **foreach** node $v \in G$ **do**
12   compute $HCE(v)$ using (6);
13   $HCE[v] \leftarrow$ compute $HCE_N(v)$ using (7);
14 **foreach** node $v \in G$ **do**
15   $NC[v] \leftarrow$ compute $NC_N(v)$ using (8), (9), (10) in turn;
16 **foreach** node $v \in G$ **do**
17   $DSCHI\_VALUE[v] \leftarrow$ compute $DSCHI(v)$ using (11);
18 **return** $DSCHI\_VALUE$;

---

*B. Cost-Effective Mutual-Influence-based Voting Approach*

Inspired by real-world voting scenarios and based on the aforementioned DSCHI method, this section introduces the CEMIV method for seed nodes selection. First, we discuss the influence score measurement of a node by taking both its own importance and the importance of its neighbor nodes into account. Then, the mutual voting mechanism and the LSUS for optimizing the selecting of seed nodes are presented. Finally, the algorithm of CEMIV is presented as a whole.

*1) Influence Score Measurement:* In real-world voting scenarios, when the influential people vote for a person, a large number of other people will also vote for the same person. In contrast, few people pay attention to the person who is voted for by the people with no influence. For this reason, we argue that the influence of each vote cast by nodes with different importance is also different. We define an index called Single Vote Score (SVS) to differentiate and quantified the efficacy of the votes cast by nodes with different importance. For a node $v$ in the network, its Single Vote Score (SVS) can be defined as:

$$SVS(v) = DSCHI(v) \quad (12)$$

It is a fact that two people usually can influence each other. The people who is more influential is more likely to influence other people, such as the sports stars, the politicians and others. Inspired by this, we define an index called Mutual Influence (MI) to measure and quantify the extent to which the nodes influence their neighbor nodes. In reality, when people try to influence others, the effect of the influence of the influencer on the person being influenced is conditioned by the importance of both the influencer and the person being influenced. Therefore, we argue that the influence of a node $v$ to its neighbor node $u$ depends on two aspects: the importance of the node $v$ itself and the sum of the importance of node $v$ and node $u$. Then, for the node $v$ and node $u$ belong to the same edge, the MI of node $v$ to node $u$ can be calculated as follow:

$$MI(vu) = \frac{\beta^{3*DSCHI(v)}}{\beta^{DSCHI(v)+DSCHI(u)}}, \beta > 1 \quad (13)$$

where $DSCHI(\cdot)$ denotes the DSCHI value of node $v$ or $u$, and the integer $\beta$ is a control coefficient used to distinguish the mutual influence of different neighbor nodes on the target node more accurately when the importance of the target node is low.

It is natural that vote holders are most likely to cast different number of votes according to the importance of the people being voted for, for example, they probably cast the maximum number of votes to the most influential person and conversely, they probably cast the least number of votes to the least influential person. At the same time, the influence of people being voted for on vote holders is also relative. If the vote holders are influential, even though the people being voted for is also influential, the vote holders will not cast a large number of votes. What's more, the influential vote holders may keep cautious during each round of voting because the influential vote holders know their action would influence other vote holders, so the number of votes cast by the influential vote holders will not large. Obviously, the vote holders with relatively low influence do not have this limitation. Inspired by this, we argue that the number of votes cast by node $u$ for its neighbor node $v$ depends on the degree of influence of node $v$ to node $u$, i.e., the MI of node $v$ to node $u$, and we define the Voting Number (VN) of node $u$ to its neighbor node $v$ as:

$$VN(uv) = \frac{\beta^{3*DSCHI(v)}}{\beta^{DSCHI(v)+DSCHI(u)}}, \beta > 1 \quad (14)$$

With the use of the newly-introduced index of MI, we synthesize the importance of the voter node itself, the importance of the nodes being voted for and the proportion of importance that this voted node holds among all neighbor nodes of the voter node to differentiate the number of votes cast by node $v$ for its neighbor nodes.

During the voting process, every node receives the votes from its neighbor nodes, then the total score every node receives is calculated. For node $v$, we define the total score it gets in each round of voting as follows:

$$Score(v) = \sum_{u \in Neighbor(v)} SVS(u) * VN(uv) \quad (15)$$

where $Neighbor(v)$ denotes the set of neighbor nodes of node $v$. After calculating the score of all nodes, we select the node with the highest score value and add it to the set of seed nodes. Then, we set the SVS value and Score value of this selected node to 0, i.e., keep this node out of the follow-up voting process.

*2) Suppression of Overlapping Influence:* According to the Three Degrees of Influence principle in networks [50], the influence of a node spreads only within the range of its third-order neighbor nodes and gradually decays with distance. Inspired by the work in Liu *et al.*[28], we here take the suppression measure for both the first-order neighbor nodes and the second-order neighbor nodes of a selected seed node, i.e., we directly reduce the score the first-order neighbor nodes, the second-order neighbor nodes, and the third-order neighbor nodes of a selected seed node can received in each round of voting. By adopting this measure, we suppress the problem of influence overlapping between different seed nodes. In the $T^{th}$ round of voting process, for the selected seed node $s$, the updated Single Vote Score $SVS(n_{s1})_T$ of its first-order neighbor node $n_{s1}$ is computed as:

$$SVS(n_{s1})_T = SVS(n_{s1})_{T-1} * \mu^2 * (\mu - 0.1), \mu \in (0.1,1] \quad (16)$$

and the updated $SVS(n_{s2})_T$ of its second-order neighbor node $n_{s2}$ is computed as:

$$SVS(n_{s2})_T = SVS(n_{s2})_{T-1} * \mu^2, \mu \in (0.1,1] \quad (17)$$

where $\mu$ is an adjustment index used to control the suppression degree of SVS value of the first-order neighbor nodes and the second-order neighbor nodes of the seed node $s$. Since the first-order neighbor nodes are more influenced by the seed node, we take stronger suppression measure for the first-order neighbor nodes and relatively weaker suppression measure for the second-order neighbor nodes.

*3) Lazy Score Updating Strategy:* In each round of voting, after selecting seed node and taking the suppression measure, only the scores of neighbor nodes within three hops of the selected node change. Liu *et al.*[28] consequently only updated the scores of neighbor nodes within three hops of the selected node. In this way, the time and space overhead of selecting seed nodes is greatly reduced. In view of the above, a strategy called LSUS is designed to further reduce the redundant calculations.

It is evident that the scores of neighbor nodes within three hops of the selected node in the $(T+1)^{th}$ round of voting are decreased compared with those in the $T^{th}$ round of voting. The LSUS is described as follows: If the node with the second highest score in the $T^{th}$ round is not in the range of three hops of the selected node, we don't need to update the scores of neighbor nodes within three hops of the selected node immediately, and instead we directly choose the node with the second highest score in the $T^{th}$ round as the seed node of the $(T+1)^{th}$ round. Meanwhile, we store the neighbor nodes within three hops of the selected node in the $T^{th}$ round. In the following rounds, once the node with the second highest score is in the range of three hops of the selected node of the current round, we then update the scores of the stored nodes.

The proposed LSUS can also be extended to other voting-based methods [27-29] to reduce the redundant calculations while achieving the same results. For a voting-based method that takes the suppression measure for the neighbor nodes within the $n^{th}$-order of a selected seed node, if the node with the second highest score in the $T^{th}$ round is not in the range of $(n+1)$ hops of the selected node, again, we don't need to update the scores of neighbor nodes within $(n+1)$ hops of the selected node in this round, on the contrary, we can directly choose the node with the second highest score in the $T^{th}$ round as the seed node of the $(T+1)^{th}$ round. Meanwhile, we store the neighbor nodes within $(n+1)$ hops of the selected node in the $T^{th}$ round. In the following rounds, once the node with the second highest score is in the range of $(n+1)$ hops of the selected node of the current round, we then update the scores of the stored nodes.

*4) Method details:* Based on the analysis above, below we present the algorithmic representation of the proposed CEMIV method.

| Algorithm 2: CEMIV Algorithm Framework |
|---|
| **Input**: $G(V,E)$, A Dictionary $DSCHI\_VALUE$ containing the $DSCHI$ value of each node |
| **Output**: Seed nodes list $S$ |
| 1     $S = \emptyset$; |
| 2     **foreach** $node\ v \in G$ **do** |

```
3     SVS[v] = DSCHI_VALUE[v];
4     foreach node v ∈ G do
5     │ s = 0;
6     │ foreach u ∈ Neighbor(v) do
7     │ │ s += β^(3*DSCHI(v)) / β^(DSCHI(v)+DSCHI(u)) * SVS[u];
8     │ Score[v] = s;
9     while |S| < K do
10    │ v_max, v_second max ← search index in Score;
11    │ Add node v_max to list S;
12    │ Score[v_max] = 0;
13    │ SVS[v_max] = 0;
14    │ Initialize list H;
15    │ foreach n_s1 ∈ first order neighbor(v_max) do
16    │ │ SVS[n_s1] = SVS[n_s1] * μ² * (μ − 0.1);
17    │ │ H ← n_s1;
18    │ foreach n_s2 ∈ first order neighbor(v_max) do
19    │ │ SVS[n_s2] = SVS[n_s2] * μ²;
20    │ │ H ← n_s2;
21    │ H ← third order neighbor(v_max);
22    │ U_memory ← H;
23    │ if v_second max not in H:
24    │   continue;
25    │ else:
26    │   foreach u ∈ U_memory do
27    │   │ s = 0
28    │   │ foreach w ∈ neighbor(u) do
29    │   │ │ s += β^(3*DSCHI(u)) / β^(DSCHI(w)+DSCHI(u)) * SVS[w];
30    │   │ Score[u] = s;
31    │   Clear list U_memory;
32    return S;
```

### C. An overall Algorithmic Framework of the Proposed CECHMV Approach

By considering all above aspects, the outline of the proposed CECHMV approach is presented as Algorithm 3. The proposed approach can strike a sound balance between time complexity and the accuracy of influential nodes identification, which is demonstrated in Section IV.

---
**Algorithm 3: CECHMV Algorithm Framework**

**Input**: $G(V, E)$
**Output**: Seed nodes list $S$
```
1   C ← Use Leiden Algorithm to detect communities;
2   foreach community C_l ∈ C do
3   │ NN(C_l) ← compute the number of nodes of each community;
4   foreach node v ∈ G do
5   │ construct the graph G_c of different communities;
6   │ foreach node l ∈ G_c do
7   │ │ compute ne_{C_l};
8   foreach node i ∈ G_c do
9   │ compute CC_N(i) using (2), (3), (4) in turn;
10  │ compute CI(i) using (5);
11  foreach node v ∈ G do
12  │ compute HCE(v) using (6);
13  │ HCE[v] ← compute HCE_N(v) using (7);
14  foreach node v ∈ G do
15  │ NC[v] ← compute NC_N(v) using (8), (9), (10) in turn;
16  foreach node v ∈ G do
17  │ DSCHI_VALUE[v] ← compute DSCHI(v) using (11);
18  S = ∅;
19  foreach node v ∈ G do
20  │ SVS[v] = DSCHI_VALUE[v];
21  foreach node v ∈ G do
22  │ s = 0;
23  │ foreach u ∈ Neighbor(v) do
24  │ │ s += β^(3*DSCHI(v)) / β^(DSCHI(v)+DSCHI(u)) * SVS[u];
25  │ Score[v] = s;
26  while |S| < K do
27  │ v_max, v_second max ← search index in Score;
28  │ Add node v_max to list S;
29  │ Score[v_max] = 0;
30  │ SVS[v_max] = 0;
31  │ Initialize list H;
32  │ foreach n_s1 ∈ first order neighbor(v_max) do
33  │ │ SVS[n_s1] = SVS[n_s1] * μ² * (μ − 0.1);
34  │ │ H ← n_s1;
35  │ foreach n_s2 ∈ first order neighbor(v_max) do
36  │ │ SVS[n_s2] = SVS[n_s2] * μ²;
37  │ │ H ← n_s2;
38  │ H ← third order neighbor(v_max);
39  │ U_memory ← H;
40  │ if v_second max not in H:
41  │   continue;
42  │ else:
43  │   foreach u ∈ U_memory do
44  │   │ s = 0
45  │   │ foreach w ∈ neighbor(u) do
46  │   │ │ s += β^(3*DSCHI(u)) / β^(DSCHI(w)+DSCHI(u)) * SVS[w];
47  │   Score[u] = s;
48  │   Clear list U_memory;
49  return S;
```

## IV. EXPERIMENTS

In this section, we report on a series of experiments to verify the performance of the proposed methods. In Section IV-A, we briefly present the experimental evaluation criteria. Then we describe the basic experimental setup and the ten real-world complex network datasets used for the experiments in Section IV-B. Finally, the results and analysis of the experiments are demonstrated in Section IV-C.

### A. Experimental Evaluation Criteria

*1) Final Infected Scale:* The most intuitive indicator to measure the influence of the seed node set is the number of nodes influenced by the seed nodes. In the SIR model, this indicator can be quantified by using the final infected scale $F(t_c)$ when the propagation reaches stability.

$$F(t_c) = \frac{n_{R_{(t_c)}}}{n} \quad (18)$$

where $n_{R_{(t_c)}}$ denotes the number of nodes that are in State R when the propagation reaches stability, $t_c$ represents the moment when the propagation reaches stability, and $n$ denotes the total number of nodes.

*2) Infected Scale $F(t)$ at time $t$:* Another important indicator to evaluate the influence of the seed node set is how quickly it can influence other nodes in the network during the propagation process. In SIR model, this indicator can be quantified by using the infected scale $F(t)$ at time $t$, which is calculated as follows:

$$F(t) = \frac{n_{I(t)} + n_{R(t)}}{n} \qquad (19)$$

where $n_{R(t)}$ denotes the number of nodes that are in State R at time $t$, $n_{I(t)}$ denotes the number of nodes that are in State I at time $t$, and $n$ denotes the total number of nodes in the network.

*3) Running Time:* One of the most important indicators to evaluate the performance of an algorithm is the running time $T$. If the running time of the algorithm is low, it denotes that this algorithm is efficiency. When the scale of the dataset increases, if the increase of the running time of the algorithm is also reasonable, then this algorithm is scalable.

*4) Balance Index:* From Section II.A we can deduce that as the average degree increases, the influence between different nodes becomes strong and the difference between the influence of different nodes also becomes more obvious [43, 44]. Thus, when the average degree of the network increases, we argue that the importance of the accuracy of influential nodes identification also increases. Inspired by this, we develop a novel index by integrating the commonly used evaluation criteria of *Final Infected Scale* and *Running Time* to quantify and measure the performance of different methods in striking a tradeoff between time complexity and the accuracy of influential nodes identification. The index denoted by *BI* is calculated as follows:

$$BI = \log_2\left(1 + \frac{F(t_c)_i - \rho}{F(t_c)_{max} - \rho}\right) - \frac{1}{e + <D>} \log_2\left(1 + \frac{\sqrt[4]{T_i}}{\sqrt[4]{T_{max}}}\right) \qquad (20)$$

where $F(t_c)_i$ denotes the final infected scale $F(t_c)$ value of the i[th] method, $F(t_c)_{max}$ denotes the maximum final infected scale $F(t_c)$ value among all the methods, $\rho$ denotes the ratio of initial infected nodes, $<D>$ denotes the average degree of the network, $T_i$ denotes the running time value of the i[th] method, and $T_{max}$ denotes the maximum running time value among all the methods. It is worth noting that, $F(t_c)_i - \rho$ represents the ratio of nodes that are actually new infected by the i[th] method and $\frac{1}{e + <D>}$ represents the adjustment coefficient, which is used to control the weight of the running time item in different networks. The closer the *BI* value of the i[th] method is to 1, the effect of balance between running time and the accuracy of influential nodes identification achieved by the i[th] method is better.

### B. Experimental Setup and Dataset

We compare the performance of our methods with the well-known methods, including the centrality methods such as H-index [20], LSS [21], ECRM [22], LFIC [23], LGC [24], EPC [51], and DomiRank [52], the improved heuristic methods such as RLGI [26], VoteRank [27], VoteRank++ [28], and VoteRank* [29], and the community-structure-based methods such as CKS [30], CSR [34], CIFR [35], K++ Shell [53], and MV [54]. In order to verify the effectiveness of the proposed methods, we carry out experiments on ten real-world networks. The specific topological characterization of the ten networks is shown in Table I, where $N$, $M$ and $<D>$ represents the number of nodes, the number of edges, and the average degree of the network, respectively. For the SIR model used in these experiments, we set the infected rate $\lambda = \mu/\beta$, where $\beta = <D>/(<D^2> - <D>)$. In particular, the $\alpha$ in Section III-A regarding the DSCHI is set to 0.7, the $\beta$ in Section III-B regarding the MI is set to 2, and the $\mu$ in Section III-B regarding the suppression of overlapping influence is set to 0.15. The experiments have been executed on a personal computer with primary memory 16GB and 2.6-GHz Intel Core i7 processor. We conduct these experiments by utilizing Python programming language and various packages such as Networkx, igraph, etc.

TABLE I
TOPOLOGICAL CHARACTERIZATION OF NETWORK DATASETS

| Dataset | N | M | < D > |
|---|---|---|---|
| Power Grid [55] | 4941 | 6594 | 2.67 |
| Collaboration [56] | 7610 | 15751 | 4.14 |
| Lastfm-Asia [57] | 7624 | 27806 | 7.29 |
| Web-Webbase [58] | 16062 | 25593 | 3.19 |
| Ca-Condmat [59] | 21363 | 91286 | 8.55 |
| Deezer-Europe [57] | 28281 | 92752 | 6.56 |
| RO [60] | 41773 | 125826 | 6.02 |
| Tech-Gnutella [58] | 62561 | 147878 | 4.73 |
| Brack [58] | 62631 | 366559 | 11.71 |
| Fe-Tooth [58] | 78136 | 452591 | 11.58 |

### C. Experimental Results and Analysis

Fig. 2 presents the experimental results of different methods in ten real-world networks on the infected scale $F(t)$ versus time $t$. In these experiments, we set the infected rate $\lambda$ to 1.5 and the ratio of initial infected nodes $\rho$ to 0.03. To ensure the accuracy of the results, the experimental results are averaged over 100 independent runs. From Fig. 1, it can be seen that as time $t$ goes by, the infected scale $F(t)$ increases too. Under the same time $t$, the seed nodes selected by our method can infect more nodes than the seed nodes selected by other 16 comparison methods, indicating that our method outperforms the other comparison methods in terms of propagation speed in all ten real-world networks. When the propagation reaches stability, the number of nodes infected by our method is still the highest. Further, as we can see from the experimental results, the methods which have taken the suppression measures of influence overlapping perform relatively well for all ten networks.

Fig. 3 shows the experimental results of different methods with different ratios of initial infected nodes for the final infected scale $F(t_c)$. In these experiments, we set the infected rate $\lambda$ to 1.5. From Fig. 2, it can be seen that as the number of seed nodes selected increases, then the final infected scale $F(t_c)$ increases too. When the number of seed nodes is low, the difference of the final infected scale $F(t_c)$ values of different methods is small. Under different ratios of initial infected nodes, the proposed method outperforms the other 16 comparison methods, which indicates that whatever the ratio of initial infected nodes is, the proposed method is able to select the seed nodes with the most excellent propagation ability on different real-world networks. This experiment shows that the seed nodes selected by our method has the superior propagation ability over other methods for different

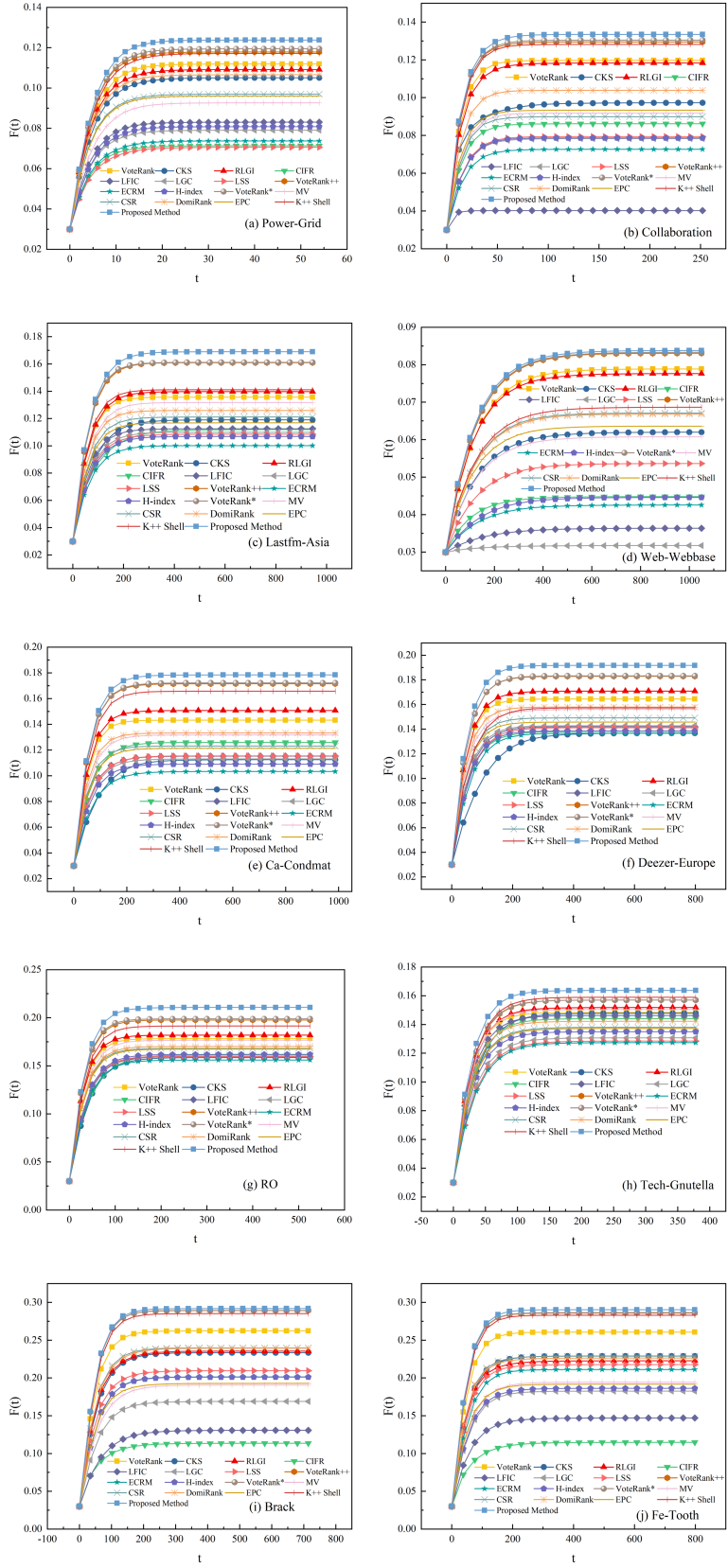

**Fig. 2.** The Infected Scale $F(t)$ versus time $t$ in ten real-world networks

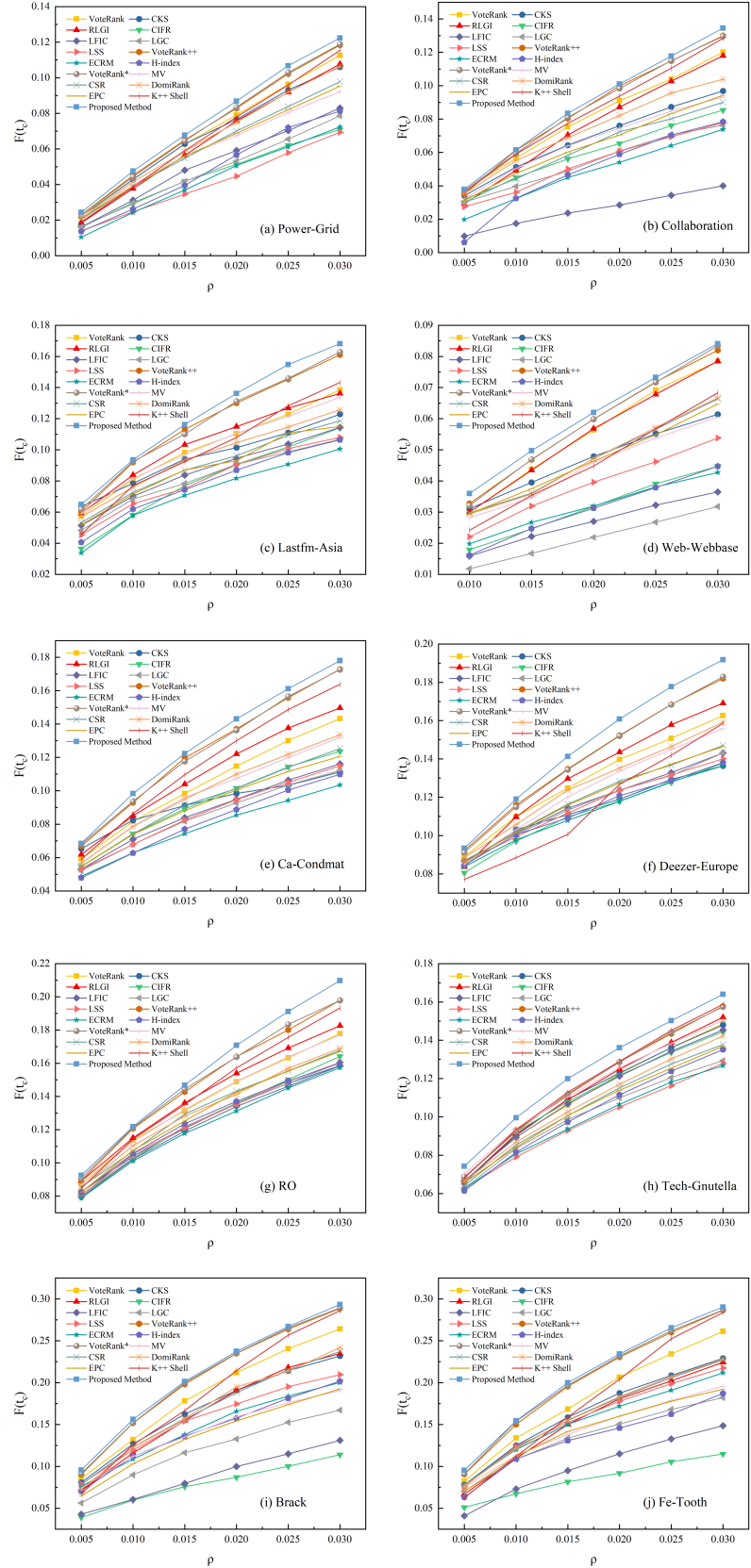

**Fig. 3.** The final infected scale $F(t_c)$ under different ratios of initial infected nodes $\rho$ in ten real-world networks

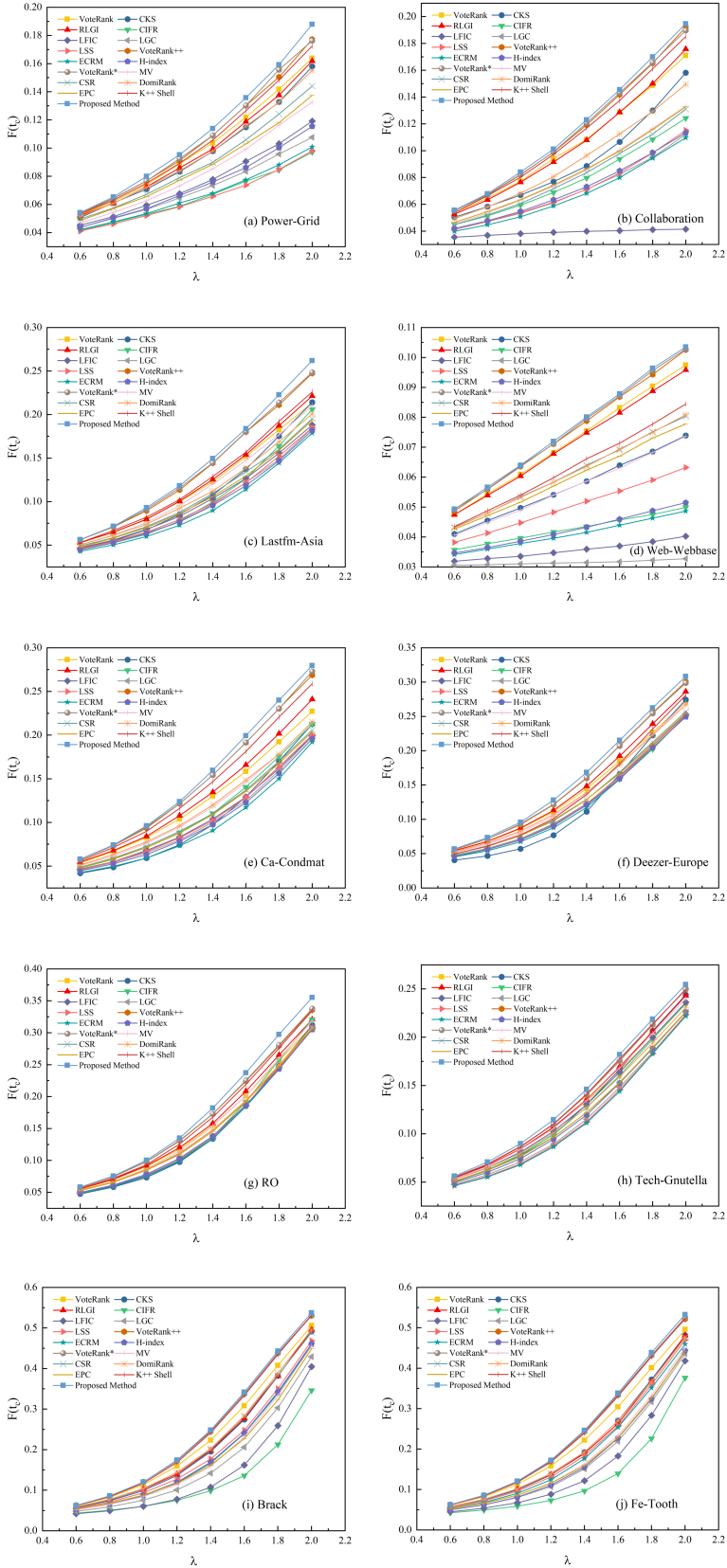

**Fig. 4.** The final infected scale $F(t_c)$ under different infected rate $\lambda$ in ten real-world networks

number of seed nodes and different types and scales of networks.

Fig. 4 illustrates the experimental results of the final infected scale $F(t_c)$ under different infected rate $\lambda$ in ten real-world networks of different methods. In these experiments, we set the ratio of initial infected nodes $\rho$ to 0.03. In general, the infected rate cannot be too small, otherwise it will block the information to spread effectively in the network no matter how to select the seed nodes. In addition, the infected rate cannot be too large, otherwise the propagation results between the seed nodes with poor propagation ability and the seed nodes with strong propagation ability are indistinguishable. In this study, we set the range of the infected rate $\lambda$ to [0.6, 2] to simulate different information dissemination scenarios. As can be seen in Fig. 3, as the selected infected rate $\lambda$ increases, the final infected scale $F(t_c)$ also increases. Under different infected rates in ten networks, the proposed method performs better in the final infected scale $F(t_c)$ than other 16 comparison methods, which indicates that the propagation ability under different infected rates of the seed nodes selected by our method is the best and the generalization ability of different information dissemination scenarios of our method is also the greatest. This experiment presents the generalization ability and the high stability of the proposed method.

Tables II and III illustrate the experimental results of the running time of different methods. The ratio $\rho$ of initial infected nodes is set to 0.03 in these experiments. From Table II and III, we can see that the running time of the proposed method is only more than that of few simplest methods, which is relatively little in most of the networks. For example, in Power-Grid network, the running time of the proposed method is only more than that of MV, LSS, H-index. As the scales of the network datasets increase, the increase of the running time of the proposed method is also reasonable, which indicates that the proposed method is scalable. On the contrary, the running time of certain methods increased irrationally as the scales of the network datasets increases.

Tables VI and V present the experimental results of the balance index *BI* of 17 methods. The ratio $\rho$ of initial infected nodes is set to 0.03 and the infected rate $\lambda$ to 1.5 in these experiments. Other experimental results of the balance index *BI* of 17 methods at different infected rate are shown in Table B.1.1- Table B.8.2 of Appendix B of the supplementary material. From Table III, we can see that the *BI* value of the proposed method is the highest in ten networks, indicating that the proposed method outperforms other comparison methods on the balance between time complexity and accuracy of influential nodes identification. Besides, compared with the method with the second highest *BI* value in the ten networks, the proposed method is improved by 6.94%, 5.23%, 6.03%, 6.60%, 5.98%, 9.29%, 7.20%, 1.54%, 1.96%, and 1.66% in Power-Grid, Collaboration, Lastfm-Asia, Web-Webbase, Ca-Condmat, Deezer-Europe, RO, Tech-Gnutella, Brack, and Fe-Tooth networks, respectively. To sum up, the proposed method can achieve the greatest performance in a relatively short running time, which indicates it can balance the time complexity and accuracy of influential nodes identification well.

TABLE II
RUNNING TIME OF NINE DIFFERENT METHODS

| Dataset | Time (s) | | | | | | | | |
|---|---|---|---|---|---|---|---|---|---|
| | CIFR | VoteRank | LGC | LSS | LFIC | RLGI | VoteRank++ | H-index | Proposed Method |
| Power-Grid | 5.75 | 1.08 | 145.86 | 0.08 | 162.96 | 0.17 | 5.99 | 0.01 | 0.17 |
| Collaboration | 10.10 | 4.00 | 222.76 | 0.22 | 135.02 | 0.25 | 19.80 | 0.18 | 0.70 |
| Lastfm-Asia | 74.74 | 6.85 | 537.17 | 0.55 | 284.90 | 0.60 | 29.19 | 0.05 | 2.91 |
| Web-Webbase | 67.53 | 12.93 | 1406.63 | 0.82 | 812.93 | 0.38 | 155.26 | 0.05 | 2.94 |
| Ca-Condmat | 191.27 | 58.66 | 4102.93 | 1.65 | 2602.42 | 2.32 | 456.41 | 0.10 | 16.74 |
| Deezer-Europe | 738.81 | 90.96 | 7107.40 | 1.50 | 6902.18 | 2.81 | 953.29 | 0.13 | 18.71 |
| RO | 1144.17 | 245.55 | 14640.38 | 2.36 | 12879.69 | 4.75 | 3307.57 | 0.20 | 23.29 |
| Tech-Gnutella | 2183.66 | 420.45 | 28305.21 | 2.57 | 7530.55 | 4.24 | 12819.96 | 0.28 | 54.09 |
| Brack | 1947.95 | 585.32 | 44630.64 | 5.23 | 93480.17 | 45.20 | 13111.01 | 0.36 | 19.22 |
| Fe-Tooth | 3098.98 | 892.65 | 67082.66 | 7.95 | 67408.22 | 64.93 | 24892.28 | 1.85 | 29.16 |

TABLE III
RUNNING TIME OF THE OTHER NINE DIFFERENT METHODS

| Dataset | Time (s) | | | | | | | | |
|---|---|---|---|---|---|---|---|---|---|
| | ECRM | CKS | VoteRank* | MV | CSR | DomiRank | EPC | K++ Shell | Proposed Method |
| Power-Grid | 0.27 | 0.91 | 6.55 | 0.09 | 0.78 | 13.82 | 90.85 | 0.33 | 0.17 |
| Collaboration | 1.54 | 1.64 | 33.45 | 0.15 | 1.75 | 14.19 | 197.36 | 1.09 | 0.70 |
| Lastfm-Asia | 6.01 | 2.31 | 44.41 | 0.21 | 2.07 | 21.86 | 387.43 | 1.02 | 2.91 |
| Web-Webbase | 1.98 | 2.20 | 162.57 | 0.19 | 2.00 | 12.40 | 1038.40 | 1.18 | 2.94 |
| Ca-Condmat | 17.24 | 6.87 | 469.70 | 0.61 | 7.48 | 720.17 | 3619.31 | 3.31 | 16.74 |
| Deezer-Europe | 17.87 | 13.73 | 979.95 | 0.78 | 15.87 | 1464.75 | 5673.45 | 6.37 | 18.71 |
| RO | 20.73 | 36.17 | 3408.16 | 1.40 | 43.14 | 9157.15 | 12856.55 | 15.99 | 23.29 |
| Tech-Gnutella | 13.71 | 65.40 | 12961.58 | 1.99 | 73.79 | 12634.69 | 22642.10 | 7.08 | 54.09 |
| Brack | 46.97 | 51.59 | 12980.53 | 3.56 | 55.90 | 140.40 | 31129.68 | 10.95 | 19.22 |
| Fe-Tooth | 122.05 | 51.47 | 24670.87 | 4.48 | 63.14 | 914.14 | 48851.98 | 14.70 | 29.16 |

TABLE IV
BALANCE INDEX $BI$ OF NINE DIFFERENT METHODS
(HIGHEST VALUE OF EACH ROW ARE BOLDED)

| Dataset | BI Value | | | | | | | | |
|---|---|---|---|---|---|---|---|---|---|
| | CIFR | VoteRank | LGC | LSS | LFIC | RLGI | VoteRank++ | H-index | Proposed Method |
| Power-Grid | 0.4326 | 0.8539 | 0.4283 | 0.4756 | 0.4687 | 0.8366 | 0.8716 | 0.6126 | **0.9557** |
| Collaboration | 0.5337 | 0.8315 | 0.3836 | 0.5091 | -0.0010 | 0.8457 | 0.8769 | 0.5171 | **0.9553** |
| Lastfm-Asia | 0.6173 | 0.7949 | 0.5380 | 0.6223 | 0.5992 | 0.7984 | 0.9058 | 0.6218 | **0.9654** |
| Web-Webbase | 0.2531 | 0.8565 | -0.1229 | 0.4911 | 0.0098 | 0.8937 | 0.8605 | 0.3289 | **0.9527** |
| Ca-Condmat | 0.6581 | 0.7820 | 0.5482 | 0.6379 | 0.5800 | 0.8365 | 0.9164 | 0.6138 | **0.9711** |
| Deezer-Europe | 0.6665 | 0.8188 | 0.6572 | 0.7309 | 0.6577 | 0.8746 | 0.8820 | 0.7284 | **0.9682** |
| RO | 0.7342 | 0.8151 | 0.6723 | 0.7700 | 0.6658 | 0.8660 | 0.8650 | 0.7761 | **0.9700** |
| Tech-Gnutella | 0.8088 | 0.8568 | 0.6664 | 0.7745 | 0.7914 | 0.9136 | 0.8502 | 0.8251 | **0.9632** |
| Brack | 0.3668 | 0.8930 | 0.5443 | 0.7416 | 0.3992 | 0.8162 | 0.9392 | 0.7201 | **0.9887** |
| Fe-Tooth | 0.3683 | 0.8876 | 0.5938 | 0.7726 | 0.4714 | 0.7880 | 0.9313 | 0.6743 | **0.9864** |

TABLE V
BALANCE INDEX $BI$ OF THE OTHER NINE DIFFERENT METHODS
(HIGHEST VALUE OF EACH ROW ARE BOLDED)

| Dataset | BI Value | | | | | | | | |
|---|---|---|---|---|---|---|---|---|---|
| | ECRM | CKS | VoteRank* | MV | CSR | DomiRank | EPC | K++ Shell | Proposed Method |
| Power-Grid | 0.4959 | 0.8019 | 0.8737 | 0.7053 | 0.7321 | 0.7546 | 0.6047 | 0.8937 | **0.9557** |
| Collaboration | 0.4520 | 0.6592 | 0.8639 | 0.6510 | 0.6004 | 0.6857 | 0.5486 | 0.9079 | **0.9553** |
| Lastfm-Asia | 0.5551 | 0.7089 | 0.9105 | 0.7897 | 0.6830 | 0.7060 | 0.5974 | 0.8367 | **0.9654** |
| Web-Webbase | 0.2613 | 0.6160 | 0.8794 | 0.6215 | 0.6990 | 0.6818 | 0.5549 | 0.7343 | **0.9527** |
| Ca-Condmat | 0.5525 | 0.6078 | 0.9147 | 0.7427 | 0.6929 | 0.7015 | 0.6021 | 0.9091 | **0.9711** |
| Deezer-Europe | 0.6977 | 0.6993 | 0.8859 | 0.8150 | 0.7547 | 0.7668 | 0.6789 | 0.8190 | **0.9682** |
| RO | 0.7431 | 0.7452 | 0.8641 | 0.8467 | 0.7868 | 0.7214 | 0.7057 | 0.9048 | **0.9700** |
| Tech-Gnutella | 0.7573 | 0.8726 | 0.8481 | 0.9079 | 0.8121 | 0.7623 | 0.7174 | 0.9486 | **0.9632** |
| Brack | 0.7051 | 0.8069 | 0.9411 | 0.6792 | 0.8229 | 0.8337 | 0.6369 | 0.9697 | **0.9887** |
| Fe-Tooth | 0.7453 | 0.8039 | 0.9324 | 0.7018 | 0.7971 | 0.7866 | 0.6326 | 0.9703 | **0.9864** |

## V. Conclusion

In this study, a method called CECHMV is proposed to strike a good balance between time complexity and the accuracy of influential nodes identification. Unlike existing methods, we measure the amount of community structural information of nodes in a fine-grain level and combine the community structural information with the hierarchy structural information of nodes to measure the importance of nodes. We further design the mutual voting mechanism and lazy updating strategy to select the seed nodes effectively and efficiently. Experimental results in 10 real-world networks datasets demonstrate that our method can achieve the highest accuracy of influential nodes identification in a relatively short time. Meanwhile, since its high applicability of different types and different scales of complex networks, the application of this proposed method is also promising. In the future, a promising direction would be to extend our work to other types of networks such as dynamic networks [61], signed networks [62] and others. In addition, we will consider solving such application-level problems such as Target-Aware IM [63], Efficient Similarity-Aware IM [64], etc.